\documentclass[conference]{IEEEtran}
\IEEEoverridecommandlockouts
\usepackage{cite}
\usepackage{amsmath,amssymb,amsfonts}
\usepackage{algorithmic}
\usepackage{graphicx}
\usepackage{textcomp}
\usepackage{xcolor}
\usepackage{url}
\usepackage{booktabs}
\usepackage{placeins}

\usepackage[acronym, nopostdot, nonumberlist, nogroupskip]{glossaries}
\makenoidxglossaries

\usepackage[capitalize, noabbrev]{cleveref}
\usepackage{xcolor}
\crefname{section}{Section}{Sections}
\Crefname{section}{Section}{Sections}
\crefname{subsection}{Section}{Sections}
\Crefname{subsection}{Section}{Sections}
\crefname{equation}{Eq.}{Eqs.}
\Crefname{equation}{Eq.}{Eqs.}
\crefname{table}{Table}{Tables}
\Crefname{table}{Table}{Tables}
\crefname{figure}{Fig.}{Figs.}
\Crefname{figure}{Fig.}{Figs.}

\newacronym{sic}{SiC}{silicon carbide}
\newacronym{mlp}{MLP}{multilayer perceptron}
\newacronym{cnn}{CNN}{convolutional neural network}
\newacronym{lstm}{LSTM}{long short-term memory}
\newacronym{node}{neural ODE}{neural ordinary differential equation}
\newacronym{rul}{RUL}{remaining useful life}
\newacronym{eol}{EoL}{end of life}
\newacronym{pof}{PoF}{physics of failure}
\newacronym{aqg}{AQG~324}{ECPE Automotive Qualification Guideline 324}
\newacronym{cwl}{CWL}{cycle-window length}
\newacronym{dsf}{DSF}{down-sampling factor}
\newacronym{mae}{MAE}{mean absolute error}
\newacronym{ra}{RA}{relative accuracy}
\newacronym{rk}{RK}{Runge--Kutta}

\def\BibTeX{{\rm B\kern-.05em{\sc i\kern-.025em b}\kern-.08em
    T\kern-.1667em\lower.7ex\hbox{E}\kern-.125emX}}
\begin{document}

\title{Failure-Mechanism Transferability of Cumulative-Damage Features for Health State Estimation of SiC Power Modules\thanks{In Germany this work of Infineon Technologies AG and Infineon Dresden AG \& Co KG is funded in the frame of the Important Project of Common European Interest on Microelectronics and Communication Technologies (IPCEI ME/CT). The IPCEI ME/CT is funded by the German Federal Ministry for Economic Affairs and Energy, the Bavarian Ministry for Economic Affairs, Regional Development and Energy, the Ministry of Economic Affairs, Industry, Climate Action and Energy of the State of North Rhine-Westphalia, the Saxon State Ministry for Economic Affairs, Labour, Energy and Climate Action and the European Union within “NextGenerationEU”. 
}\thanks{\copyright~2026 IEEE. This work has been accepted for presentation at the 52nd Annual Conference of the IEEE Industrial Electronics Society (IECON~2026). Personal use of this material is permitted. Permission from IEEE must be obtained for all other uses, in any current or future media, including reprinting/republishing this material for advertising or promotional purposes, creating new collective works, for resale or redistribution to servers or lists, or reuse of any copyrighted component of this work in other works.}}

\author{%
\IEEEauthorblockN{Mattia Scarpa}
\IEEEauthorblockA{\textit{Dept.\ of Information Engineering}\\
\textit{University of Padova}\\
Padova, Italy\\
scarpamatt@dei.unipd.it}
\and
\IEEEauthorblockN{Evgeny Kusmenko}
\IEEEauthorblockA{\textit{Infineon Technologies Dresden AG \& Co.\ KG}\\
Dresden, Germany\\
Evgeny.Kusmenko@infineon.com}
\and
\IEEEauthorblockN{Francesco Toso}
\IEEEauthorblockA{\textit{Newtwen S.r.l.}\\
Padova, Italy\\
francesco.toso@newtwen.com}
\and
\IEEEauthorblockN{Mattia Bruschetta}
\IEEEauthorblockA{\textit{Dept.\ of Information Engineering}\\
\textit{University of Padova}\\
Padova, Italy\\
bruschet@dei.unipd.it}
\and
\IEEEauthorblockN{Ruggero Carli}
\IEEEauthorblockA{\textit{Dept.\ of Information Engineering}\\
\textit{University of Padova}\\
Padova, Italy\\
carlirug@dei.unipd.it}
\and
\IEEEauthorblockN{Simon Achatz}
\IEEEauthorblockA{\textit{Infineon Technologies AG}\\
Munich, Germany\\
Simon.Achatz@infineon.com}
}

\maketitle

% \todo{first sentence of abstract sounds provocative, do we have  proof?}
% This is what is typically found in literature where experiment are usually conducted only on synthetic data or specifically picked campaign. Even worse some times tests are done on a single power module and the problem of validation still remains an open topic. Mainly for the lack of dataset available and established. 
\begin{abstract}
Data-driven health-state estimators for \gls{sic} power modules
typically report their performance on a single accelerated-aging
campaign, and how that performance transfers to a different failure
mechanism is rarely tested. We benchmark five reference methods from
the prognostics and condition-monitoring literature against a
physics-informed \gls{node} on two \gls{sic} power-cycling campaigns
driven by structurally different failure mechanisms, solder-layer
fatigue and wire-bond lift-off, under a per-module $k$-fold
protocol. The
\gls{node} is evaluated under two input regimes that share the rest of
the pipeline: the baseline electrical precursors and a set of
cumulative thermoelectric features. Every reference method degrades on
the wire-bond campaign, with average errors growing and precision 
decreasing with respect to their performance on the soldered campaign.
The \gls{node} fed with the cumulative features keeps its
soldered-campaign metrics on both mechanisms, with differences inside
the fold-to-fold variance, while the same architecture fed with the
baseline precursors falls back to the reference-method cluster. The
input representation contributes at least as much as the architecture
to failure-mechanism transferability of a health-state estimator.
\end{abstract}


\section{Introduction}
\label{sec:intro}

The adoption of \gls{sic} technology in modern power converters,
and most notably in automotive traction inverters, exploits higher load levels
and elevated junction temperatures that subject the device package to substantial
thermo-mechanical stress responsible for the dominant wear-out mechanisms,
namely bond-wire lift-off and die-attach delamination~\cite{kovacevic2015}.
Tracking the health state of each module from in-service sensor measurements
is therefore crucial information for maintenance scheduling, warranty-grade
lifetime estimation, and the optimal control and performance management
of the electric vehicle. The quantity adopted as target in this work is
a normalised health indicator on the interval $[0,1]$, where $0$ corresponds
to a pristine module and $1$ to its \gls{eol}. Repeated active
power-cycling tests are the established methodology for accelerated aging
of these devices, and the \gls{aqg}~\cite{aqg324} provides
the standardised qualification procedure that frames such tests for quality
and warranty assurance. Beyond their qualification purpose, these tests
also yield datasets suitable for the analysis of degradation phenomena
and for the development of condition-monitoring models.

\subsection{Related Work}

Lifetime modeling for power semiconductor modules has historically followed the \gls{pof} tradition.
Coffin--Manson estimates combined with Miner's rule and Stress-Strain modeling for variable load profiles
~\cite{kovacevic2010,kovacevic2015,ciappa2002,miner1945,ceccarelli2019,barbagallo2021}
provide closed-form lifetime predictions but rely on assumed indicator shapes and on
parameters that vary across batch, package and mission profile.

Data-driven estimators attempt to learn a health value directly from 
the available sensor data, with no explicit fatigue law. The reference 
architectures in literature, due to lack of publicly available dataset, 
principally refer to benchmarks such as the C-MAPSS turbofan engine simulation
~\cite{saxena2008}, but the same architectural lineage is also supported by the 
broader NASA Prognostics Center of Excellence (PCoE) dataset family, which includes 
IGBT accelerated aging, lithium-ion battery, FEMTO bearing and milling data sets~\cite{nasaPCoE}.
However, these datasets remain limited in their coverage of power-electronics
aging, specifically, with none reproducing the power-cycling regime and the 
wear-out mechanisms of SiC power modules.

Several architectures have been benchmarked on these data, the most notable being
a \gls{cnn}~\cite{babu2016} and a \gls{lstm} regressor~\cite{zheng2017};
these architectures, however, inherit the limitations of the underlying datasets, which,
while representative, remain far from a realistic and generalisable test, making difficult the
validation of the resulting methods. Other studies have instead leveraged industrial datasets
that are not publicly available~\cite{olschewski2025}, achieving a more concrete validation but
still bound to a single dataset representing a single degradation mechanism, so any observed
performance ranking is conditional on the specific campaign on which it was obtained.

In parallel, considerable effort has been devoted to identifying which physical quantities
are the most informative degradation precursors and which combinations of them constitute
reliable health indicators. Several reviews~\cite{hanif2019,ni2020,kumar2026} cover hundreds
of contributions on this question for power electronic devices; for \gls{sic} MOSFETs, the on-state
voltage drop $V_{\mathrm{DS,on}}$, the on-state resistance $R_{\mathrm{DS,on}}$ and the junction
temperature $T_j$ have emerged as the most consistently reported indicators~\cite{diNuzzo2022,villalobos2025}.
In the \gls{pof} tradition, per-cycle thermal stress is aggregated via Miner's rule
into a scalar cumulative damage value used as the lifetime indicator itself.
This quantity has not, to our knowledge, been adopted as an input feature by
data-driven health-state estimators, which still rely on raw or filtered
electrical waveforms.

A small number of recent contributions has begun to address the input-representation question explicitly.
Parametric fits using $V_{\mathrm{DS,on}}$, $T_j$ and their multivariate combination have been compared
on a \gls{sic} campaign~\cite{diNuzzo2023}, with feature-set-dependent variations in fit quality of comparable
magnitude to the differences between architectures. The influence of window length and downsampling on
snapshot \gls{mlp} regressors has been examined in~\cite{achatz2026}, with the same
observation that performance is more sensitive to the chosen feature configuration than to network depth
or width. The broader use of physics-informed in condition monitoring and power electronics domains is reviewed in~\cite{fassi2024,scarpa2025}.
Despite these advances, the comparisons remain confined to a single dataset each, and the transferability
of any feature configuration across datasets with different load profiles, packaging variability and
acquisition conditions is still an open question.
\subsection{Contribution}

In this work we investigate the role of the input representation in health-state estimation for \gls{sic}
power modules across two power-cycling campaigns that share the \gls{aqg} qualification framework~\cite{aqg324}
but exercise two structurally different modules and, consequently, two distinct dominant wear-out mechanisms.
The first dataset originates from the campaign of~\cite{olschewski2025}, performed on modules with conventional
solder die-attach, where lifetime is governed by progressive solder-layer fatigue that translates into a smooth,
quasi-exponential drift of the on-state voltage. The second dataset, introduced in our earlier work~\cite{scarpa2026phm},
exercises modules realised with a sintered packaging die-attach: the suppression of the solder layer removes the
dominant fatigue channel of the first dataset, and lifetime becomes governed instead by wire-bond lift-off, a
mechanism that produces non-monotonic perturbations and multi-regime degradation traces on the same electrical
observables. The two datasets therefore probe the same estimation problem on the same kind of device but under
qualitatively different physics, so a fair comparison across them is also a stress test of how much of a method's
published performance is mechanism-specific and how much transfers across failure modes.

On this basis we benchmark, under a per-module $k$-fold cross-validation protocol with a common
set of metrics, two complementary lineages of data-driven estimators. The first comprises established
architectures of the prognostics literature, so far validated on public benchmarks of adjacent domains:
a \gls{cnn}~\cite{babu2016} and an \gls{lstm} regressor~\cite{zheng2017}, whose transferability
to \gls{sic} power-cycling data is itself an open question. The second comprises methods recently proposed
for \gls{sic} condition monitoring and validated on a single proprietary campaign each: the parametric fit
of~\cite{diNuzzo2023}, the snapshot \gls{mlp} of~\cite{achatz2026} and the patch-based Transformer used
in~\cite{olschewski2025}, whose dataset-specific performance has not been cross-checked
against a second campaign. Alongside these we evaluate a physics-informed \gls{node}~\cite{chen2018node}
drawn from our earlier work~\cite{scarpa2026phm}.
The
physics-informed model is evaluated under two input regimes, the baseline electrical signal
precursors $V_{\mathrm{DS,on}}$ and $I_{\mathrm{load}}$, and a set of cumulative thermoelectric
features that encode the accumulated thermal and electrical stress experienced by the component,
so that the contribution of the representation is separated from that of the architecture.
The drop in accuracy across the two failure mechanisms varies
considerably with architecture and input representation. The \gls{node}
evaluated on the cumulative features achieves both the highest
accuracy on each campaign and the smallest such drop.

The remainder of the paper is organised as follows. \Cref{sec:modules} introduces the two
power modules, the corresponding datasets, the dominant degradation phenomena and the \gls{eol}
criterion adopted as health-state target. \Cref{sec:methods} describes the methodologies compared
in the study, including the cumulative feature set fed to the physics-informed model. \Cref{sec:results}
reports the comparative evaluation under the unified protocol.
By unified protocol we mean that the two campaigns are
brought onto a common evaluation basis, rather than compared through the
figures published by their respective sources: both datasets are
acquired on the same class of power-cycling test bench and with the same
measurement chain for the electrical and thermal quantities, they are
reduced to the same feature set, and every method is trained, validated
and scored with the same code, the same health-state target and the same
four metrics, with module-level disjointness between training and
evaluation enforced throughout.
\Cref{sec:discussion} closes
with the discussion of the results and the open questions.

\section{Modules, Degradation, and Datasets}
\label{sec:modules}

\begin{figure*}[t]
\centering
\includegraphics[width=\linewidth]{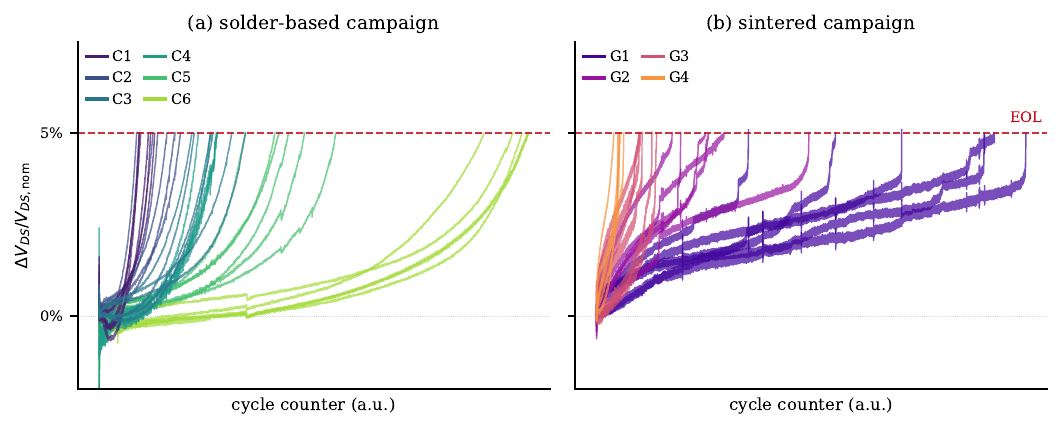}
\caption{Normalised on-state voltage drop $\Delta V_{DS,\mathrm{norm}}$ of every device
against its cycle counter for the two campaigns. (a) Solder-based dataset
of~\cite{diNuzzo2023,olschewski2025,achatz2026}: predominantly smooth quasi-exponential trajectories
driven by progressive solder fatigue. (b) Sintered dataset of~\cite{scarpa2026phm}:
continuous drift due to bond-pad crack propagation, with rapid step-like rises
corresponding to individual wire lift-off events. The horizontal dashed line marks the
$5\%$ \gls{eol} threshold; cycle-counter values are anonymised to preserve dataset
confidentiality.}
\label{fig:vds_both}
\end{figure*}

We consider two accelerated power-cycling campaigns performed at Infineon Technologies
on automotive \gls{sic} half-bridge inverter modules with different die-attach
technologies, which define two qualitatively distinct dominant wear-out mechanisms.
Both campaigns are run within the \gls{aqg} qualification framework~\cite{aqg324};
\cref{fig:vds_both} reports the normalised on-state voltage drop
$\Delta V_{DS,\mathrm{norm}}$ of every device against its cycle counter for the two
campaigns.

\subsection{Solder-Based Campaign}
\label{sec:dataset_solder}

The first dataset, originally used in~\cite{diNuzzo2023} and subsequently
in~\cite{olschewski2025,achatz2026}, comprises power-cycling runs on conventional
solder die-attach modules across the six
stress conditions reported in~\cref{tab:cond_solder}. The conditions combine two heating
regimes, $\mathrm{PC}_\mathrm{min}$ ($t_\mathrm{heat}=60$~s) and $\mathrm{PC}_\mathrm{sec}$
($t_\mathrm{heat}=1.5$~s), with three junction-temperature swing levels
$\Delta T_j \in \{80, 100, 120\}~\mathrm{K}$. Each condition collects eight to twelve transistors,
for a total of $60$ degradation paths that reach the $5\%$ on-state voltage \gls{eol}
criterion. After median filtering of the per-cycle measurements, the trajectories
reported in~\cref{fig:vds_both}(a) follow predominantly smooth quasi-exponential profiles
characteristic of progressive die-attach fatigue~\cite{kovacevic2015}.

\begin{table}[t]
\centering
\caption{Operating conditions of the solder-based campaign~\cite{diNuzzo2023,olschewski2025,achatz2026}.}
\label{tab:cond_solder}
\begin{tabular}{cccccc}
\toprule
ID & Regime & $t_\mathrm{heat}$ [s] & $t_\mathrm{cool}$ [s] & $I_\mathrm{load}$ [A] & $\Delta T_j$ [K] \\
\midrule
C1 & $\mathrm{PC}_\mathrm{min}$ & 60  & 60  & 267 & 120 \\
C2 & $\mathrm{PC}_\mathrm{min}$ & 60  & 60  & 254 & 100 \\
C3 & $\mathrm{PC}_\mathrm{min}$ & 60  & 60  & 229 & 80  \\
C4 & $\mathrm{PC}_\mathrm{sec}$ & 1.5 & 3.5 & 283 & 120 \\
C5 & $\mathrm{PC}_\mathrm{sec}$ & 1.5 & 3.5 & 261 & 100 \\
C6 & $\mathrm{PC}_\mathrm{sec}$ & 1.5 & 3.5 & 232 & 80  \\
\bottomrule
\end{tabular}
\end{table}

\subsection{Sintered Campaign}
\label{sec:dataset_sintered}

The second dataset, introduced in our earlier work~\cite{scarpa2026phm}, comprises
power-cycling runs on automotive \gls{sic} half-bridge modules whose sintered packaging
suppresses solder fatigue at the chip--substrate interface and leaves
wire-bond fatigue at the chip-side bond pads as the dominant package-level mechanism.
Twenty-four modules are distributed across the four operating-condition groups reported
in~\cref{tab:cond_sintered} (six modules per group), spanning two heating-pulse durations
($t_\mathrm{heat} \in \{1.5, 30\}~\mathrm{s}$) and four load-current levels
($I_\mathrm{load} \in [670, 775]~\mathrm{A}$); the inlet coolant temperature is adjusted
per group to keep the junction-temperature peak close to $175\,^\circ\mathrm{C}$ across
conditions. The same $5\%$ on-state voltage \gls{eol} criterion is adopted and is again
the primary event observed. As shown in~\cref{fig:vds_both}(b), most trajectories drift
smoothly under continuous crack propagation in the bond-pad metallization, with rapid
step-like rises corresponding to wire lift-off events visible on several of the
modules.

\begin{table}[t]
\centering
\caption{Operating conditions of the sintered campaign~\cite{scarpa2026phm}; each group
contains $N=6$ modules and the junction-temperature peak is kept close to
$T_{v_j,\max}\approx 175\,^\circ\mathrm{C}$ across groups.}
\label{tab:cond_sintered}
\begin{tabular}{ccccc}
\toprule
ID & $t_\mathrm{heat}$ [s] & $t_\mathrm{cool}$ [s] & $I_\mathrm{load}$ [A] & $T_\mathrm{inlet}$ [$^\circ$C] \\
\midrule
G1 & 1.5 & 3.5 & 695 & 75 \\
G2 & 1.5 & 3.5 & 750 & 55 \\
G3 & 30  & 30  & 670 & 75 \\
G4 & 30  & 30  & 775 & 35 \\
\bottomrule
\end{tabular}
\end{table}

\subsection{End-of-Life Criterion}
\label{sec:eol}

For both campaigns the \gls{eol} event is defined by the \gls{aqg} qualification
threshold of a $5\%$ relative increase in the on-state voltage drop~\cite{aqg324}:
\begin{equation}
\frac{\Delta V_{DS}(n)}{V_{DS,\mathrm{nom}}} \geq 5\%,
\label{eq:eol_criterion}
\end{equation}
where $\Delta V_{DS}(n)$ is the per-cycle drift with respect to the value $V_{DS,\mathrm{nom}}$
measured on the pristine module. The first cycle at which~\cref{eq:eol_criterion} is
crossed defines the lifetime $N_{\mathrm{EoL}}$ of each module and grounds the
health-state target adopted in~\cref{sec:methods}.

\section{Methodologies}
\label{sec:methods}

We compare five reference methods proposed in the literature against the
physics-informed \gls{node} drawn from our earlier work, evaluated on
both datasets of \cref{sec:modules} under a common target, protocol and
set of metrics. The reference methods are grouped, for clarity, in three
blocks of increasing modelling capacity: a parametric power-law fit
specific to \gls{sic} modules, two established deep regressors from the
prognostics literature, and two recent deep methods proposed
specifically for \gls{sic} condition monitoring.

The common health-state target adopted by all methods is the relative
lifetime
\begin{equation}
y_n \;=\; \frac{n}{N_{\mathrm{EoL}}}\;\in\;[0,\,1],
\label{eq:target}
\end{equation}
where $n$ is the cycle index and $N_{\mathrm{EoL}}$ is the first cycle at
which the on-state voltage rise reaches the \gls{aqg} threshold of
$5\%$ above the pristine value. Four metrics are reported on the
evaluation cycles, all aggregated as mean and standard deviation across
the splits: the \gls{mae} $\mathrm{MAE}=\mathbb{E}|\hat{y}-y|$, the
coefficient of determination $R^2$, the Saxena $\alpha$-accuracy with
$\alpha = 0.2$~\cite{saxena2010} that scores the fraction of predictions
inside a lifetime-shrinking $\alpha$-cone, and the \gls{ra} on the late
part of the lifetime~\cite{saxena2010}. The two datasets are evaluated
independently, each with the split protocol declared by its own source.
On the first dataset we follow the stratified random split
of~\cite{olschewski2025,achatz2026}, with a $75/25$ partition inside
each cycling batch repeated for five random seeds whose metrics are
then averaged. On the second dataset we follow the per-module
four-fold leave-one-condition-out split of~\cite{scarpa2026phm}, in
which one of the four cycling batches is held out at each fold while
the remaining three form the training pool. Both protocols enforce
module-level disjointness between training and evaluation.

\subsection{Parametric Fit}
\label{ssec:dinuzzo}

The parametric fit of~\cite{diNuzzo2023} treats the two \gls{sic}
degradation indicators reported as informative for solder fatigue, the
on-state voltage rise $\Delta V_{\mathrm{DS,on}}$ and the maximum
junction temperature rise $\Delta T_{j,\max}$, as power laws of the
relative lifetime,
\begin{equation}
f(y) \;=\; \alpha_0 \;+\; \alpha_1\,y^{\alpha_2},
\label{eq:dinuzzo}
\end{equation}
with the three coefficients estimated from the training trajectories by
nonlinear least squares pooled per cycling batch. At inference time the
lifetime is recovered by analytical inversion of \cref{eq:dinuzzo} on
the observed indicator; we reproduce the three input variants reported
in the original paper, namely $V_{\mathrm{DS,on}}$-only,
$T_{j,\max}$-only, and the multivariate combination obtained as the
average of the two single-indicator inversions. The method has no
trainable architecture beyond the three coefficients per indicator and
is included as a lightweight reference of the parametric tradition for
\gls{sic} modules.

\subsection{Established Deep Baselines}
\label{ssec:cnn-lstm}

Two deep regressors that have become established references in the
prognostics literature, originally validated on the C-MAPSS turbofan
benchmark~\cite{saxena2008}, are included as the convolutional and
recurrent reference architectures of the comparison. We adopt the \gls{cnn} regressor of~\cite{babu2016},
later refined in~\cite{li2018}, which stacks two one-dimensional
convolutional blocks with ReLU activations followed by a dense head
whose final layer produces a scalar lifetime estimate; each block
aggregates the temporal context inside the input window.
The \gls{lstm} regressor of~\cite{zheng2017} replaces the convolutional
stack with two recurrent layers feeding the same-style dense head; the
recurrent state acts as an implicit memory of the input trajectory.
Both networks operate on the same sliding window of multi-channel
measurements, and in our experiments are fed with the
$V_{\mathrm{DS,on}}$ and $I_\mathrm{load}$ inputs used by the other
deep models, so that the only design choice not shared with the rest
of the comparison is the architecture itself.

\subsection{SiC-Specific Deep Methods}
\label{ssec:achatz-patchtst}

Two recent methods proposed specifically for \gls{sic} condition
monitoring complete the reference set. The snapshot \gls{mlp}
of~\cite{achatz2026} maps a recent input window directly to the
lifetime target of \cref{eq:target} through a fully connected network
of uniform hidden width and ReLU activations: the input window of
length \gls{cwl} is downsampled by a factor \gls{dsf}, flattened, and
passed to the network, with the number of layers and the layer width
selected by an Optuna hyperparameter search on the training subset
within the ranges declared in the original paper. We adopt the
timeseries configuration $\mathrm{CWL}{=}600$, $\mathrm{DSF}{=}10$ and
the multivariate input $\{\Delta V_{\mathrm{DS,on}},\,
\Delta R_{\mathrm{th}},\,\Delta I_\mathrm{D}\}$ reported by the authors
as the best operating point of the architecture.

The Transformer of~\cite{olschewski2025} approaches lifetime estimation
indirectly: a PatchTST model~\cite{nie2023patchtst}, which tokenises the
input series into overlapping patches and processes them through a
transformer encoder, forecasts the next segment of
$\Delta V_{\mathrm{DS,on}}$ from a window of past observations; the
lifetime is recovered by iterating the forecast forward until the
predicted indicator crosses the \gls{eol} threshold. We replicate the
configuration declared as best in the original paper, namely
$\mathrm{CWL}{=}1800$, $\mathrm{DSF}{=}20$, single-channel
$\Delta V_{\mathrm{DS,on}}$ input and a $3{:}1$ input-to-forecast
length ratio, together with the patch size, the model dimension, the
number of attention heads and the number of encoder layers reported by
the authors, and apply the threshold-crossing inference at evaluation
time.

\begin{table}[t]
\centering
\footnotesize
\setlength{\tabcolsep}{4pt}
\caption{Model complexity and per-cycle inference cost. FLOPs are reported per new cycle observation under stateful deployment: window-based methods run one forward pass over their input window, the \gls{node} variants advance their latent state by a single integration step.}
\label{tab:complexity}
\begin{tabular}{@{}llrrl@{}}
\toprule
\textbf{Method} & \textbf{Type} & \textbf{Params} & \textbf{FLOPs/cycle} & \textbf{Input/cycle} \\
\midrule
Parametric~\cite{diNuzzo2023}   & Closed-form & $\sim 6$    & $\sim\!10$ (negl.)        & $(1,)$    \\
DeepCNN~\cite{babu2016}         & CNN         & $1.3\,$k    & $6.1\!\times\!10^{4}$     & $(30,3)$  \\
DeepLSTM~\cite{zheng2017}       & LSTM        & $1.6\,$k    & $7.4\!\times\!10^{4}$     & $(30,3)$  \\
SnapshotMLP~\cite{achatz2026}   & MLP         & $6.2\,$k    & $1.2\!\times\!10^{4}$     & $(180,)$  \\
PatchTST~\cite{nie2023patchtst} & Transformer & $6.9\,$k    & $5.4\!\times\!10^{4}$     & $(90,1)$  \\
\midrule
NODE-base                       & Neural ODE  & $1.2\,$k    & $2.4\!\times\!10^{3}$     & $(2,)$    \\
NODE-cum                        & Neural ODE  & $1.3\,$k    & $2.6\!\times\!10^{3}$     & $(6,)$    \\
\bottomrule
\end{tabular}
\end{table}

\begin{table*}[!t]
\centering
\caption{Validation metrics across the two campaigns. Values are mean$\,\pm\,$std across folds, reported in percent of the unit lifetime range ($\times 100$); best per column in \textbf{bold}.}
\label{tab:results}
\setlength{\tabcolsep}{4.5pt}
\renewcommand{\arraystretch}{1.10}
\footnotesize
\begin{tabular}{l cccc | cccc}
\toprule
 & \multicolumn{4}{c|}{Soldered campaign} & \multicolumn{4}{c}{Sintered campaign} \\
\cmidrule(lr){2-5} \cmidrule(lr){6-9}
Method & MAE$\,\downarrow$ & $R^{2}\,\uparrow$ & $\alpha$-acc$\,\uparrow$ & RA$\,\uparrow$
       & MAE$\,\downarrow$ & $R^{2}\,\uparrow$ & $\alpha$-acc$\,\uparrow$ & RA$\,\uparrow$ \\
\midrule
Parametric   & $12.1\pm1.9$ & $70.3\pm10.9$ & $38.4\pm5.3$ & $61.9\pm3.9$ & $16.9\pm7.2$ & $40.6\pm38.5$ & $40.6\pm16.6$ & $62.1\pm14.5$ \\
DeepCNN      & $9.9\pm2.5$ & $77.7\pm13.9$ & $47.1\pm5.6$ & $74.7\pm4.1$ & $16.0\pm2.0$ & $43.4\pm19.6$ & $38.0\pm1.6$ & $65.8\pm3.4$ \\
DeepLSTM     & $9.4\pm2.6$ & $78.9\pm13.7$ & $52.1\pm4.5$ & $75.5\pm4.9$ & $13.5\pm2.2$ & $64.7\pm10.1$ & $34.4\pm3.2$ & $68.5\pm3.8$ \\
SnapshotMLP  & $4.8\pm0.6$ & $94.2\pm1.7$ & $68.8\pm5.7$ & $\mathbf{86.2\pm1.3}$ & $15.3\pm2.8$ & $56.0\pm15.1$ & $34.5\pm5.3$ & $63.1\pm7.0$ \\
PatchTST     & $4.6\pm0.3$ & $90.4\pm4.1$ & $65.7\pm7.4$ & $81.0\pm4.0$ & $34.3\pm1.1$ & $-94.3\pm9.7$ & $5.3\pm0.8$ & $-12.2\pm3.2$ \\
NODE-base    & $7.3\pm0.8$ & $88.3\pm2.2$ & $65.4\pm8.0$ & $76.0\pm2.9$ & $7.7\pm1.0$ & $85.0\pm2.6$ & $49.3\pm8.1$ & $71.7\pm1.8$ \\
NODE-cum     & $\mathbf{4.1\pm0.9}$ & $\mathbf{95.7\pm1.8}$ & $\mathbf{73.9\pm5.0}$ & $85.9\pm3.0$ & $\mathbf{4.9\pm1.9}$ & $\mathbf{94.7\pm2.9}$ & $\mathbf{79.2\pm12.6}$ & $\mathbf{85.9\pm4.5}$ \\
\bottomrule
\end{tabular}
\end{table*}

\subsection{Physics-Informed Neural ODE}
\label{ssec:node}

The physics-informed model drawn from our earlier
work~\cite{scarpa2026phm} is a \gls{node}~\cite{chen2018node} in which a
continuous-time hidden state $h(t)$ evolves under a learnable vector
field $g_\theta$ driven by the input,
\begin{equation}
\frac{\mathrm{d}h}{\mathrm{d}t} \;=\; g_\theta\!\bigl(h(t),\,x(t)\bigr),
\qquad
\hat{y}(t) \;=\; \pi_\theta\!\bigl(h(t)\bigr),
\label{eq:node}
\end{equation}
with $\pi_\theta$ a small projection head mapping the hidden state to
the scalar lifetime estimate; the integration is carried out on the
cycle grid by an explicit \gls{rk} step. The same architecture is
evaluated under two input regimes that share the rest of the pipeline.
The first uses the baseline electrical precursors
$V_{\mathrm{DS,on}}$ and $I_\mathrm{load}$ that are also fed to the
convolutional and recurrent baselines of \cref{ssec:cnn-lstm}; the
second uses the cumulative thermoelectric feature set
$\{S_{T_j},\;S_{\Delta T_j},\;S_I,\;D_{\mathrm{Miner}}\}$ introduced
in~\cite{scarpa2026phm}, in which each feature is the running integral
of one electrothermal quantity from the start of life up to the current
cycle $n$,
\begin{equation}
\begin{aligned}
S_{T_j}(n)        &\;=\; \sum_{i \le n} T_{j,i},
& S_{\Delta T_j}(n) &\;=\; \sum_{i \le n} \Delta T_{j,i}, \\
S_{I}(n)          &\;=\; \sum_{i \le n} I_{i},
& D_{\mathrm{Miner}}(n) &\;=\; \sum_{i \le n} \frac{n_i}{N_{f,i}},
\end{aligned}
\label{eq:cumfeat}
\end{equation}
where $D_{\mathrm{Miner}}$ implements the rule of cumulative damage
of~\cite{miner1945} on the cycle-by-cycle Norris--Landzberg stress
profile~\cite{norris1969}, so that the feature set encodes the history
of thermal, electrical and damage accumulation observed by the module
up to cycle $n$. The
contrast between the two input regimes isolates the contribution of the
representation under a fixed architecture and provides the feature
ablation reported in \cref{sec:results}.

The seven methods of this section also differ in their
inference budget. \Cref{tab:complexity} reports, for each one, the
trainable parameter count and the FLOPs required to process a single
new cycle observation in stateful deployment: window-based methods run
one forward pass over their input window, while the \gls{node} variants
advance their latent state by a single integration step. All neural
methods sit within a single order of magnitude in parameter count
($\sim 1$--$7\,$k), while per-cycle FLOPs span two orders of magnitude
across the comparison.

\begin{figure*}[t]
\centering
\includegraphics[width=\linewidth]{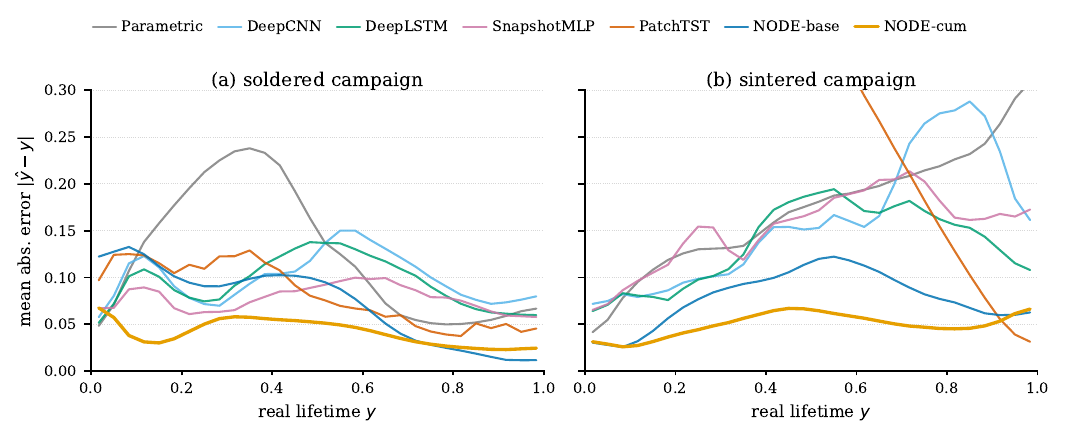}
\caption{Mean absolute prediction error $|\hat{y}-y|$ as a function of
the real lifetime $y$ on (a)~the soldered campaign and (b)~the sintered
campaign, averaged over all validation modules of every fold and binned
with $30$ bins of width $\approx 0.033$ smoothed by a three-bin rolling
mean. NODE-cum is the only method that retains a comparable
error-versus-lifetime profile across the two campaigns; PatchTST on the
sintered campaign diverges past $y\approx 0.6$, consistent with the
failure of the forecast-then-threshold paradigm on multi-regime
trajectories.}
\label{fig:cross_overview}
\end{figure*}

\section{Results}
\label{sec:results}

\begin{figure}[!t]
\centering
\includegraphics[width=\linewidth]{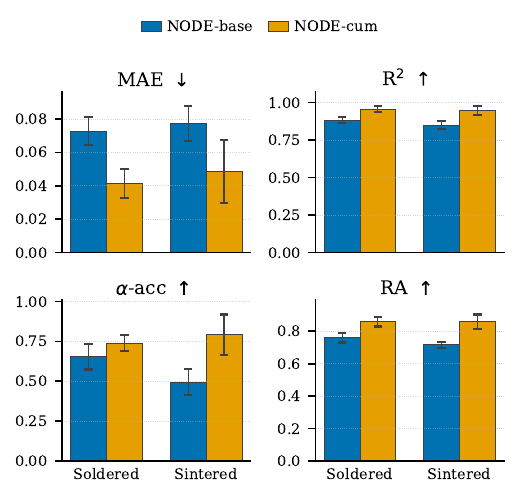}
\caption{Feature-set ablation on the \gls{node} architecture: same model
trained under the baseline electrical inputs (NODE-base) and under the
cumulative thermoelectric inputs (NODE-cum) of~\cref{ssec:node}. The
four panels report MAE, $R^2$, $\alpha$-accuracy and RA on the two
campaigns.}
\label{fig:node_ablation}
\end{figure}

The aggregated metrics of \cref{tab:results}, the error-versus-lifetime
profile of \cref{fig:cross_overview} and the feature-set ablation of
\cref{fig:node_ablation} expose three complementary views of the
cross-campaign behaviour of the seven methods of~\cref{sec:methods};
all numerical entries are expressed in percent of the unit lifetime
range. Differences smaller than one fold-to-fold standard deviation are
not claimed as significant; the central object of comparison is the
cross-campaign drop of each method between the soldered and sintered
metrics.

On the soldered campaign the parametric fit, the SnapshotMLP and the
PatchTST forecaster match the metrics reported in their original
papers~\cite{diNuzzo2023,achatz2026,olschewski2025}. The DeepCNN and
the DeepLSTM stay just below $10\%$ MAE and around $78\%$ $R^{2}$, in
line with the normalised RMSE these models report on C-MAPSS
in~\cite{babu2016,zheng2017,saxena2008}. However, their $\alpha$-accuracy
drops to around $50\%$, against the $65$--$74\%$ of the SnapshotMLP,
the PatchTST and the two \gls{node} configurations. The average error
matches the C-MAPSS validation, but the prediction band widens close
to \gls{eol}, where the Saxena $\alpha$-cone gets
tighter~\cite{saxena2010} and the lifetime estimate matters most for
maintenance scheduling and for performance and usage control.
NODE-cum has the lowest MAE and the highest $R^{2}$ and
$\alpha$-accuracy, and stays within one standard deviation of the
SnapshotMLP on RA. Overall, the SnapshotMLP, the PatchTST and the two
\gls{node} configurations sit in the upper half of the table on
$R^{2}$, $\alpha$-accuracy and RA. The DeepCNN, the DeepLSTM and the
parametric fit stay in the lower half, with about half of their
late-life predictions outside the $\alpha$-cone.
\Cref{fig:cross_overview}(a) shows the per-lifetime error profile of
each method: NODE-cum stays flat around $4\%$, the parametric fit has
the mid-life hump typical of a power-law fit, and the four deep
regressors trace overlapping curves within one standard deviation of
each other. Close to \gls{eol} the DeepCNN and the DeepLSTM stay above
the rest, consistent with their lower $\alpha$-accuracy.

Moving to the sintered campaign every method deteriorates. The
parametric fit drops to about $17\%$ MAE and $41\%$ $R^{2}$, a clear
worsening from an already weak baseline on the soldered side. The
DeepCNN, the DeepLSTM and the SnapshotMLP settle in a similar region
with MAE between $13\%$ and $16\%$, $R^{2}$ between $43\%$ and $65\%$,
and $\alpha$-accuracy between $34\%$ and $38\%$; $R^{2}$ and
$\alpha$-accuracy take the largest hits, in some cases close to
halving the soldered values, and RA degrades along the same line. The
fold-to-fold variance grows on every metric, pointing to a difficulty
both in capturing the degradation phenomenon and in generalising the
little structure these architectures manage to learn from it. The
worst case is the PatchTST forecaster, with MAE around $34\%$ and
$R^{2}$ well below zero. \Cref{fig:cross_overview}(b) shows the
collapse concentrating past $y\approx 0.6$, where the iterative
$\Delta V_{\mathrm{DS,on}}$ extrapolation must cover the longest
horizon: the wire-bond lift-off events that drive the sintered
degradation produce step-like rises in the trajectory whose timing
and amplitude are stochastic and cannot be anticipated from the past
window, breaking the predictable-shape assumption that the
forecast-then-threshold paradigm of~\cref{ssec:achatz-patchtst} relies
on and making this approach intrinsically fragile on this type of
trajectory. The two \gls{node} configurations
behave differently. NODE-base degrades only marginally and stays close
to its own soldered metrics across folds, although without reaching
the level of the best methods on the soldered side. NODE-cum holds the
same level of performance as the best entries of the soldered
campaign, with only a slight variance increase that is most visible on
$\alpha$-accuracy and that is consistent with the higher stochasticity
of the sintered phenomenon.

It is also worth noting that the DeepLSTM holds $R^{2}$ noticeably
better than the DeepCNN and the SnapshotMLP on the sintered campaign.
Together with the relative stability of the two \gls{node}
configurations, this suggests an advantage of architectures that
carry an intrinsic memory of the input history.

The feature ablation reported in \cref{fig:node_ablation} isolates the
contribution of the input representation under a fixed architecture,
since NODE-base and NODE-cum share identical layers, optimisation
schedule and split protocol of~\cref{ssec:node}, and differ only in
the input set. Under the baseline electrical precursors the \gls{node}
stays within the band of the SnapshotMLP and the PatchTST forecaster on
the soldered campaign and noticeably ahead of every other reference
method on the sintered campaign, an effect attributable to the
continuous-time inductive bias of the architecture
in~\cref{eq:node}. Switching the input to the cumulative thermoelectric
set of~\cref{eq:cumfeat} produces a further reduction in MAE of
approximately $40\%$ relative on both campaigns and lifts $R^{2}$ above
$94\%$ in either case. NODE-cum gives nearly the same metrics on both
campaigns, with the differences between soldered and sintered values
falling inside the fold-to-fold variance, and the cross-campaign drop
in performance observed on every other method of \cref{tab:results}
does not occur for this configuration. The same
\gls{node} architecture moves from sitting at the same level as the
SnapshotMLP on the soldered campaign to clearly ahead of every other
entry on the sintered campaign once the cumulative features are
introduced, supporting the claim that the representation contributes
at least as much as the architecture to failure-mechanism
transferability.

\section{Discussion and Conclusion}
\label{sec:discussion}

The two campaigns considered in this work are representative of two
distinct degradation phenomena. On the soldered modules the wear-out
is dominated by the fatigue of the chip-substrate solder layer, and
the resulting on-state voltage drift is essentially monotone and
regular cycle after cycle. On the sintered modules that layer is
suppressed and the dominant mechanism becomes the wire-bond lift-off
at the chip-side bond pads, which proceeds through stochastic
step-like rises rather than a smooth drift. The two regimes are
clearly distinguishable in the raw trajectories of \cref{fig:vds_both}
and in the per-lifetime error profiles of
\cref{fig:cross_overview}(a),(b), and the results of
\cref{sec:results} indicate that on the second regime even the
reference methods proposed and validated on \gls{sic} modules struggle
to keep their soldered-campaign performance without some form of
prior information about the cumulative stress history.

On this evidence, the input representation contributes at least as
much as the architecture to failure-mechanism transferability. The
PatchTST forecaster of~\cite{olschewski2025}, which models the
indicator trajectory, and the SnapshotMLP of~\cite{achatz2026}, which
regresses the lifetime ratio directly, are structurally different and
differ by an order of magnitude in parameter count, yet they degrade
by comparable margins between the two campaigns. The same \gls{node}
architecture, conversely, moves from sitting at the same level as the
SnapshotMLP on the soldered campaign to clearly ahead of every other
entry on the sintered campaign once its input is exchanged for the
cumulative thermoelectric set of~\cref{eq:cumfeat}, while the
underlying network, optimisation schedule and split protocol are kept
identical. The transferability margin obtained by NODE-cum should be
read against the choice of input set: the cumulative features build
on a thermo-mechanical history prior, and the Norris--Landzberg model
enters through the per-cycle damage contribution to
$D_{\mathrm{Miner}}$. Their effectiveness on aging modalities that
do not involve accumulated thermo-mechanical stress is an open
question.

It is also worth noting that the two campaigns originate
from two distinct products, manufactured through different processes and
driven by different degradation phenomena. Despite these substantial
differences, the same approach carries over unchanged, with the same
architecture, training procedure, protocol and input set. Methods that
instead rely on careful, hand-crafted modelling incur an unnecessary
burden, since a new engineering effort would be required for every new
module and, potentially, for every manufacturing variation.

We have benchmarked five reference methods from the prognostics and
condition-monitoring literature against a physics-informed \gls{node}
on two \gls{sic} power-cycling campaigns that share the qualification
threshold of~\cref{eq:eol_criterion} but exercise structurally
different package technologies. The \gls{node} fed with the cumulative
thermoelectric features attains the lowest error on each campaign and
the smallest cross-campaign degradation in the comparison, supporting
the central claim that the input representation contributes at least
as much as the architecture to failure-mechanism transferability of a
health-state estimator.
This advantage comes at no significant deployment
cost: as reported in \cref{tab:complexity}, the per-cycle inference
budget of the \gls{node} sits in the low kFLOPs range, below every
window baseline of the comparison. The combination of stateful update,
$\mathcal{O}(\text{kFLOPs})$ per-cycle cost and a hidden state of a few
dozen floats is compatible with execution on automotive-grade
microcontrollers.

A remark is due on the thermal quantities that feed the
cumulative features. The junction temperature is not measured directly:
it is reconstructed on the test bench through the thermo-sensitive
electrical parameter methodology, with a calibration fitted per module,
so that $T_j$ and $\Delta T_j$ are virtual quantities derived from
electrical measurements rather than sensor readings. On a qualification
bench this reconstruction is accurate, but it rests on a calibration
phase that a converter in the field does not have, so the online
availability of the junction temperature is the enabling condition for
the cumulative features. That condition is no longer out of reach.
Digital-twin virtual thermal sensing now delivers the internal thermal
state of power converters and electric drives in real time, from
physics-based models identified once and then executed on board, with
experimental validation on industrial
hardware~\cite{torchio2025tpel,torchio2025tte}. The cumulative features
are therefore not confined to the test bench: they become deployable as
soon as such a twin runs alongside the estimator, on the same class of
embedded platform the \gls{node} already targets.

A natural follow-up is to test the same methods under operating
regimes that depart from the constant-cycle accelerated protocol of
\gls{aqg}, in particular under variable mission profiles closer to
in-service automotive use, where the ability of cumulative-stress
features and continuous-time architectures to track non-stationary
loads can be examined directly.

\end{document}